\documentclass[twocolumn,trackchanges]{aastex63}

\usepackage[super]{nth}
\usepackage{savesym}
\savesymbol{tablenum}
\usepackage[T1]{fontenc}
\usepackage{inputenc}
\usepackage[range-units = brackets,tophrase={-},seperr,repeatunits=false]{siunitx}
\restoresymbol{SIX}{tablenum}

\graphicspath{{./}{figures/}}

\accepted{\today}

\submitjournal{ApJ}

\shorttitle{AGN FEEDBACK IN NGC~5044}
\shortauthors{Schellenberger et al.}

\begin{document}

\title{A new feedback cycle in the archetypal cooling flow group NGC~5044}
\correspondingauthor{Gerrit Schellenberger}
\email{gerrit.schellenberger@cfa.harvard.edu}

\author[0000-0002-4962-0740]{Gerrit Schellenberger}
\affiliation{Center for Astrophysics $|$ Harvard \& Smithsonian, 60 Garden St., Cambridge, MA 02138, USA}

\author{Laurence P. David}
\affiliation{Center for Astrophysics $|$ Harvard \& Smithsonian, 60 Garden St., Cambridge, MA 02138, USA}

\author{Jan Vrtilek}
\affiliation{Center for Astrophysics $|$ Harvard \& Smithsonian, 60 Garden St., Cambridge, MA 02138, USA}

\author[0000-0002-5671-6900]{Ewan O'Sullivan}
\affiliation{Center for Astrophysics $|$ Harvard \& Smithsonian, 60 Garden St., Cambridge, MA 02138, USA}

\author[0000-0002-1634-9886]{Simona Giacintucci}
\affiliation{Naval Research Laboratory, 4555 Overlook Avenue SW, Code 7213, Washington, DC 20375, USA}

\author[0000-0002-9478-1682]{William Forman}
\affiliation{Center for Astrophysics $|$ Harvard \& Smithsonian, 60 Garden St., Cambridge, MA 02138, USA}

\author{Christine Jones}
\affiliation{Center for Astrophysics $|$ Harvard \& Smithsonian, 60 Garden St., Cambridge, MA 02138, USA}

\author[0000-0002-8476-6307]{Tiziana Venturi}
\affiliation{INAF - Istituto di Radioastronomia, via Gobetti 101, I-40129 Bologna, Italy}

\begin{abstract}
The fate of cooling gas in the centers of galaxy clusters and groups is still not well understood, as is also the case for the complex process of triggering active galactic nucleus (AGN) outbursts in their central dominant galaxies, and the consequent re-heating of the gas by the AGN jets. With the largest known reservoir of cold molecular gas of any group-dominant galaxy and three epochs of AGN activity visible as cavities in its hot gas, NGC~5044 is an ideal system in which to study the cooling/AGN feedback cycle at the group scale. We present  VLBA observations of NGC~5044 to ascertain the current state of the central AGN. We find a compact core and two small jets aligned almost in the plane of the sky, and in the orthogonal direction to the location of cavities. We construct the radio/sub-mm spectral energy distribution (SED) over more than three orders of magnitude. We find that below 5\,GHz the spectrum is best fit by a self-absorbed continuous injection model representing emission coming from the jets, while the higher frequencies show clear signs of an advection dominated accretion flow. We derive a black hole mass and accretion rate consistent with independent measurements. We conclude that the age of the jets is much younger than the innermost cavities, marking the start of a new feedback cycle.
\end{abstract}


\section{Introduction} \label{sec:intro}
X-ray observations of the centers of relaxed galaxy clusters and groups  show that these systems contain large amounts of hot X-ray emitting gas that should be radiatively cooling on timescales less than a Hubble time (see e.g., \citealp{Fabian1984-kj,Voit2011-ko,Stern2019-ph}). 
The primary uncertainty in the original cooling flow scenario was the ultimate fate of the cooling gas, and the amount of energy input. 
While  diffuse  H$\alpha$ emission  was  commonly  found  within  the central  dominant  galaxy  (CDG)  in  cooling  flows  (\citealp{Heckman1981-mu,Hu1985-aw}) and star formation was observed in cluster cooling flows as early as \citealp{McNamara1989-vo}, the observed star formation rates were orders of magnitude less than the inferred mass deposition rates of the hot gas (e.g.,  \citealp{David2001-il,McNamara2007-xa}). 

\subsection{Cold gas in cooling flow clusters}
Only at the turn of the millennium did the first hints of a robust solution to the problem begin to appear with the comparison of high angular resolution X-ray and radio observations (e.g., \citealp{Bohringer1993-iv,Churazov2000-uy}). These analyses showed that supermassive black holes (SMBH) at the centers of X-ray bright atmospheres were radiatively faint, but mechanically powerful, and major modifications to the cooling flow scenario were implemented after AGN-induced cavities and shocks within cooling flows had been discovered (e.g., \citealp{McNamara2000-pl,Blanton2003-os,Fabian2003-te,Forman2007-ew,Randall2010-gc,Werner2018-xb}). 
Cavities provide a relatively simple method for estimating the mechanical power of jets, since the energy required to evacuate the hot gas can be estimated from the cavity volume and the surrounding pressure; combining 
the energy with a measure of the timescale then yields an averaged jet mechanical power.
The surrounding gas gets heated in the turbulent wake of the buoyantly rising cavities. 
It has been shown that the mechanical power of the AGN is sufficient to prevent the bulk of the hot gas from cooling in many systems (\citealp{Birzan2004-pw,Dunn2006-py,OSullivan2011-xb}), leaving small quantities to fuel the limited star formation seen in many cooling-flow CDGs (e.g., \citealp{Allen1995-zk,Rafferty2006-qf,Donahue2007-pk,Quillen2008-ap}).  
The star formation rates derived from Spitzer and Herschel data correlate with the radiative cooling times of the hot gas (\citealp{Egami2006-ez,Rawle2012-wd}), confirming the connection.

CO surveys over the past decade have
shown that a substantial fraction of CDGs in cluster
cooling flows harbor detectable quantities of molecular gas
(\citealp{Edge2001-mi,Salome2003-ge,OSullivan2018-yk,Babyk2019-cn,Olivares2019-rs}).  Warm molecular  gas  has  also  been  detected  in  CDGs  by  emission from  vibrationally  excited  molecular  hydrogen  (e.g., \citealp{Jaffe1997-wb,Donahue2000-hy,Egami2006-qt}). 

Although large amounts of molecular gas have been found in galaxy clusters, galaxy groups offer an even more interesting view on the feedback cycle triggered by the central AGN.
In nearby galaxy groups (distance $< \SI{50}{Mpc}$) the cooling process can be studied in great detail by resolving the individual giant molecular clouds, rather than just tracing molecular filaments as in typical galaxy clusters.
Moreover, the shorter cooling times compared to those of clusters  mean we can sometimes observe evidence of multiple cycles of AGN activity (e.g., cavities) visible in the intra cluster medium (ICM).
The impact of feedback by the AGN is also stronger in galaxy groups due to their shallower gravitational potential with respect to clusters. 

\subsection{NGC~5044}
\begin{figure*}[htb]
    \centering
    \includegraphics[width=0.99\textwidth]{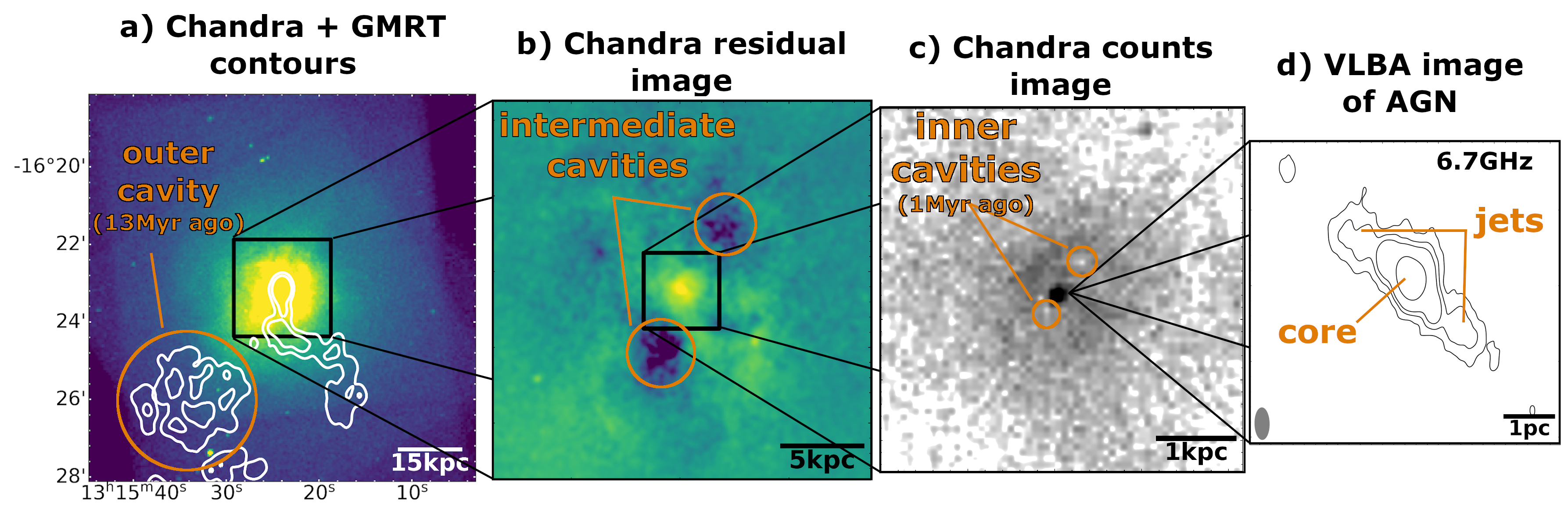}
    \caption{Cavities at various distances from the AGN related to the oldest outburst (a; adapted from \citealp{Giacintucci2011-ru}), intermediate outbursts (b; adapted from \citealp{David2009-hn}), and the most recent cavities (c; adapted from \citealp{David2017-ig}). The right panel (d) shows the high resolution radio image of the AGN (see Section \ref{ch:results}). }
    \label{fig:cavities}
\end{figure*}

NGC~5044 is one of the X-ray brightest groups in the sky. Since it is very nearby, it offers the possibility to study the AGN feedback cycle in great detail.
A wealth of multifrequency data is available for NGC~5044 making it an ideal object for studying correlations between gas properties over a broad range of temperatures. 
H$\alpha$ filaments, ro-vibrational H$_2$ line emission, [CII] line emission, and CO emission show that some gas must be cooling out of the hot phase (\citealp{Kaneda2008-xs,David2014-jn,Werner2014-vw}). 
ALMA, ACA, and IRAM single
dish observations of NGC 5044 showed it to have the
largest amount of molecular gas identified among cool core galaxy
groups (e.g., \citealp{Schellenberger2020-vl}).
Prior studies using deep, high-resolution X-ray data have shown that the hot gas within the central region has been perturbed by at least three cycles of AGN outbursts and by the motion of the central galaxy within the group potential. 
Mechanical heating by AGN-inflated cavities is sufficient to offset radiative cooling of the gas within the central $\SI{10}{kpc}$. 
The outermost (and oldest) of these X-ray cavities (Fig. \ref{fig:cavities}a) is filled with a population of relativistic particles emitting synchrotron radiation traced by deep, low frequency GMRT observations. 
Several smaller, radio-quiet cavities are detected at intermediate
distance in the Chandra X-ray residual image (Fig. \ref{fig:cavities}b), indicating  another AGN outburst cycle. 
The innermost bipolar pair of cavities (Fig. \ref{fig:cavities}c) is aligned with the orientation of molecular gas and dust within the central $\SI{2}{kpc}$.
The large amount of molecular gas available to fuel the central AGN in NGC~5044 suggests that we are seeing the AGN right before (or at the start of) a new outburst phase. 

\subsection{The central radio source in NGC~5044}
The central radio continuum source in NGC~5044 has a flux of $\sim \SI{30}{mJy}$ at $\SI{1}{GHz}$ and decreases at higher frequencies implying a negative spectral index\footnote{We define the spectral index $\alpha$ as $S_\nu \propto \nu^\alpha$, where $S_\nu$ is the flux density at frequency $\nu$}. 
However, the flux at 228GHz is significantly higher ($\sim\SI{45}{mJy}$), 
which suggests the existence of a second spectral component with positive spectral index at higher frequencies. 

\cite{Schellenberger2020-vl} note a time variability of the continuum source on timescales of years at mm wavelength, implying an upper limit on the size of at least one component of the source of $\sim\SI{1}{pc}$, consistent with the ALMA data.
The authors also give an upper limit on the continuum source extent based on the high resolution ALMA observations at $\SI{244}{GHz}$: $70\times\SI{40}{mas}~(10.5\times\SI{6}{pc})$. 
They also report two absorption features in the CO(2-1) spectra of the continuum source, with an indication of a spatially variable absorption profile. The authors explain this as due to variability with possible extent of the continuum source below the ALMA resolution. 

Despite the many studies focused on this galaxy group, very little is known about the high frequency continuum source in the center of the dominant galaxy.
We present VLBA observations of the central continuum radio source to understand the most recent feedback cycle, morphology, and absorption features in the continuum source of NGC~5044.
Our detailed analysis of the SED from the low frequency radio to the sub-mm regime shows a link from cold, infalling gas to energetic outflows giving rise to feedback.

This paper is structured as follows. In Section \ref{ch:observations}, we illustrate the VLBA, SCUBA-2, and ATCA data reduction. In section \ref{ch:results}, the results of the VLBA imaging and spectral analysis, as well as the SED, and time variability considerations are presented. Section \ref{ch:discussion} discusses the results, in particular the jet direction and their energy injection, the advection dominated accretion, and links the time variability to the infall of molecular clouds. Section \ref{ch:summary} summarizes the findings. 

We adopt a systemic velocity of  $\SI{2757}{km\,s^{-1}}$ (heliocentric) for NGC~5044 and a luminosity distance of $\SI{31.2}{Mpc}$ (\citealp{Tonry2001-wi}) to be consistent with previous works (e.g., \citealp{David2017-ig,Schellenberger2020-vl}).
This results in a physical scale in the rest frame of NGC~5044 of $\SI{1}{\arcsec}=\SI{150}{pc}$. Uncertainties are given at the $1\sigma$ level throughout the paper.

\section{Observations}
\label{ch:observations}
In the following, we describe the data reduction of our recent VLBA data revealing the parsec scale structure of the AGN, and the archival JCMT and ATCA data used for the SED in Section \ref{ch:results}.

\subsection{VLBA}
Our VLBA program (NRAO project ID BS283, PI: Schellenberger) to observe the central continuum source in NGC~5044 with VLBA was carried out in March and April 2020. The observations were split into 4 parts for source visibility reasons: The C band observation was performed on March 13 and 19, and the X band observation was done on March 20 and April 13. The four observations used all 10 VLBA antennas and experienced none to very little downtime of individual antennas due to local weather conditions. We used PT (Pie Town, NM) as reference antenna.

Each band was divided into 2 sub-bands, centered on $\SI{4.932}{GHz}$ and $\SI{5.060}{GHz}$ for C band, and $\SI{8.432}{GHz}$ and $\SI{8.560}{GHz}$ for X band. Each sub-band is sampled by 256 channels of $\SI{500}{kHz}$ width, yielding a total bandwidth of $\SI{256}{MHz}$ for each C and X band. For all four sessions, we observed J0927+3902 for 5 minutes for bandpass and gain calibration, followed a loop consisting of a 10 min target scan of NGC~5044, and a 2 min phase reference scan of 1314-134. This sums to an on target time of 220 min at each band.

Data from the 10 VLBA antennas were processed using the Socorro (NM) correlator with the Roach Digital Backend (RDBE) and Digital Down Converter (DDC), and a 2 second averaging interval.

The correlator output was reduced using the NRAO Astronomical Image Processing System (AIPS; \citealp{Greisen2003-le}) Version 31DEC19 and the VLBAUTIL tools. Each of the four observations was reduced individually. First, we applied online flags distributed with the data, and flagged edge channels, and the first 20 seconds of the first scan with the UVFLG task.
We corrected the Earth orientation parameters (VLBAEOPS), and calculated corrections for the ionospheric dispersive delay (VLBATECR), followed by amplitude corrections for digital sampling errors (VLBACCOR). Before calibrating the bandpass shape (VLBABPSS) using our calibration observation of J0927+3902, we corrected for instrumental phases and delays using the pulse calibration table (VLBAPCOR).
Post-bandpass calibrations were applied to correct the autocorrleations (VLBAAMP), and the antenna parallactic angles (VLBAPANG).
Finally, we calculate solutions from the fringe fitting algorithm (VLBAFRGP), using our phase-reference source 1314-134, with the CUBE interpolation and a solution interval of 1 minute. 

\begin{figure*}[htb]
    \centering
    \includegraphics[width=0.675\textwidth]{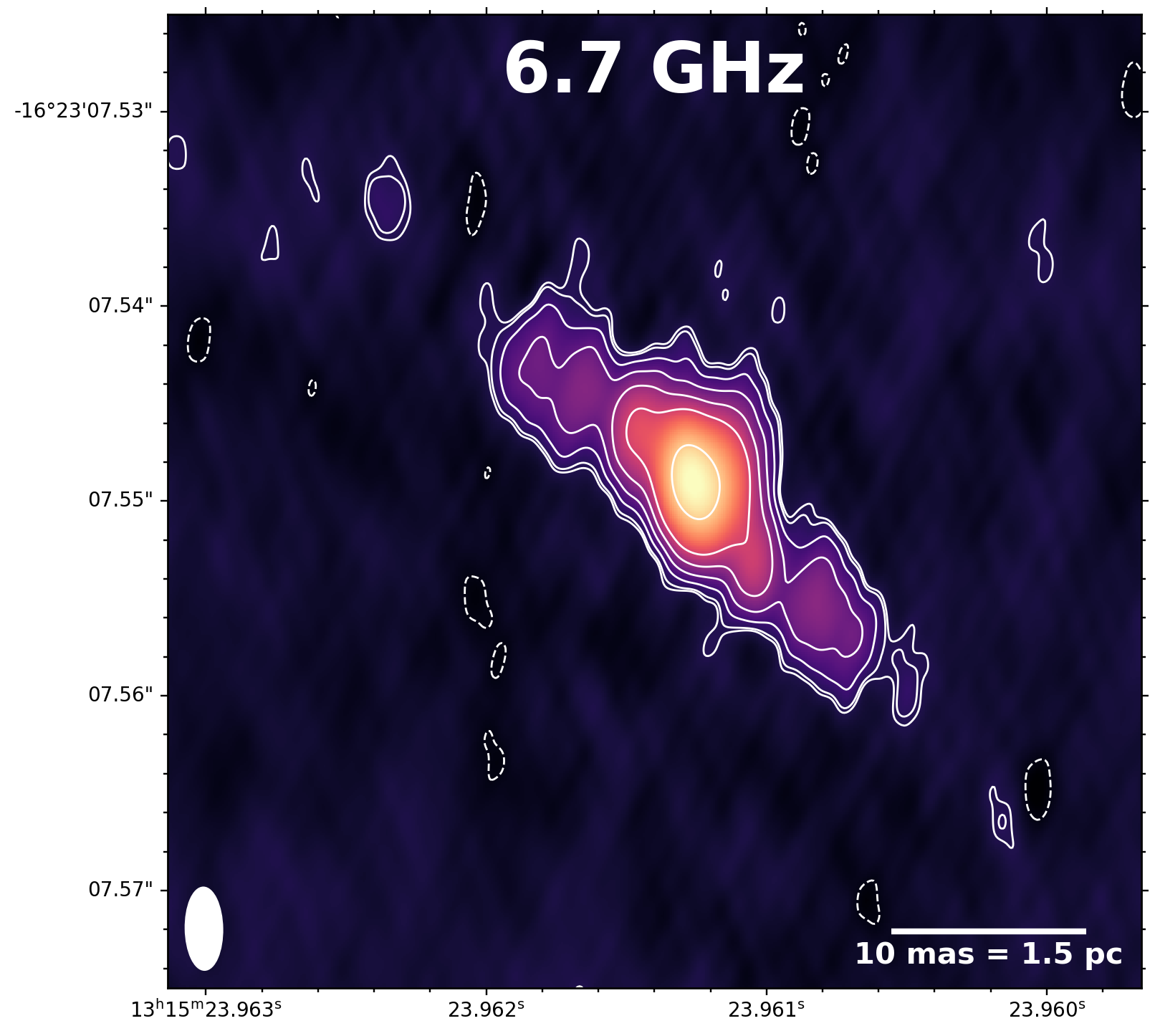}
    \includegraphics[width=0.3\textwidth]{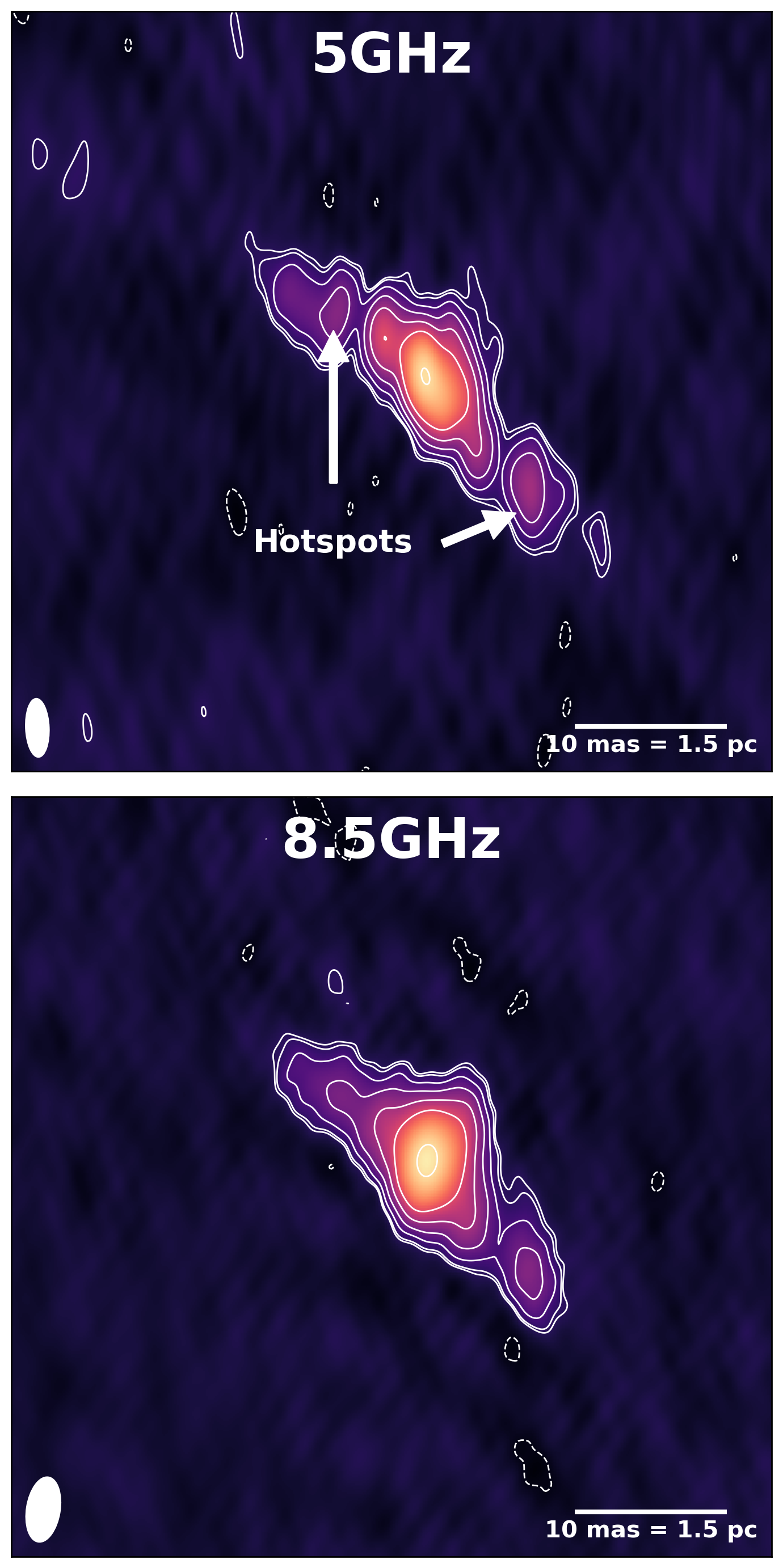}\\
    \includegraphics[width=0.45\textwidth]{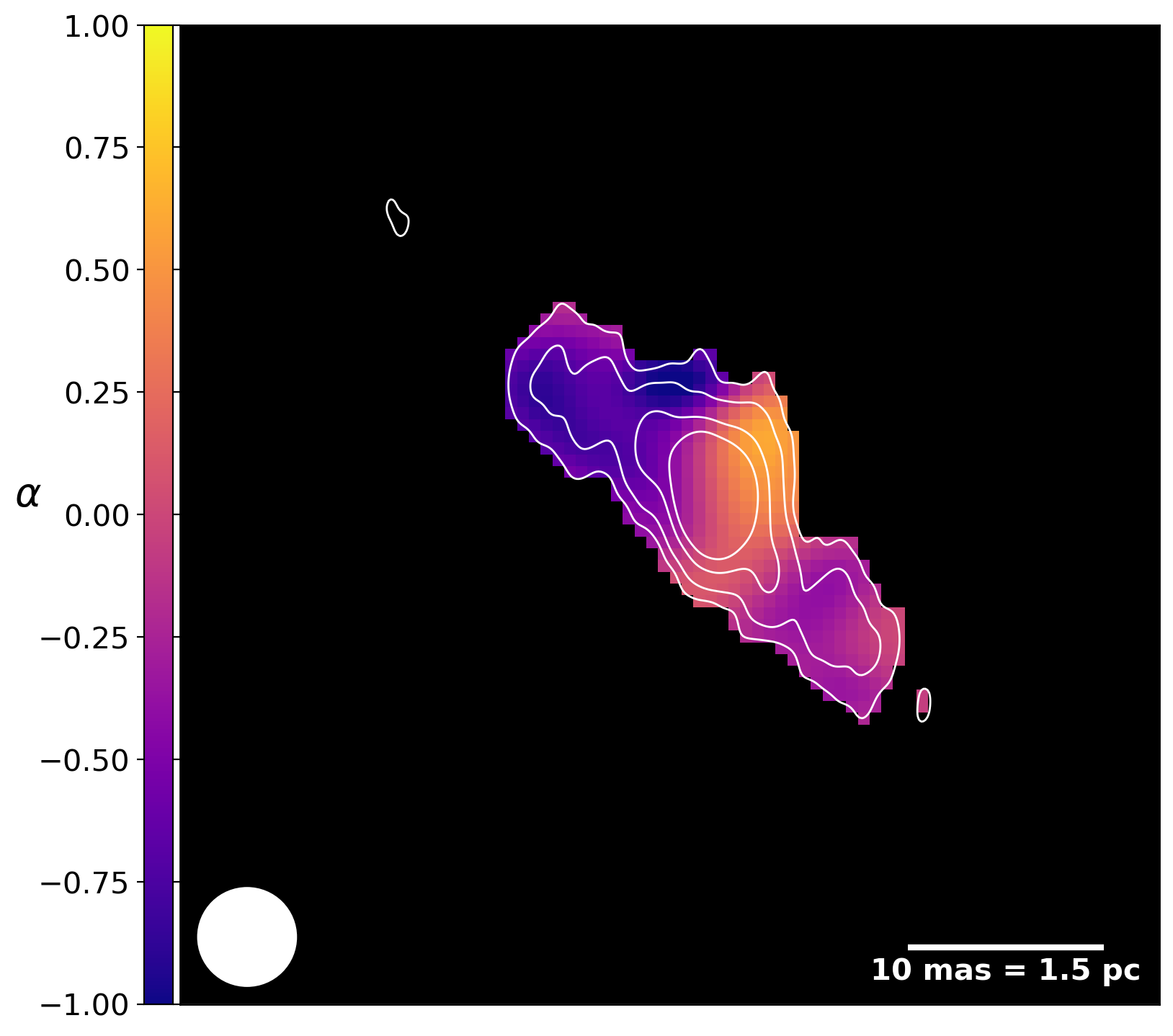}
    \includegraphics[width=0.45\textwidth]{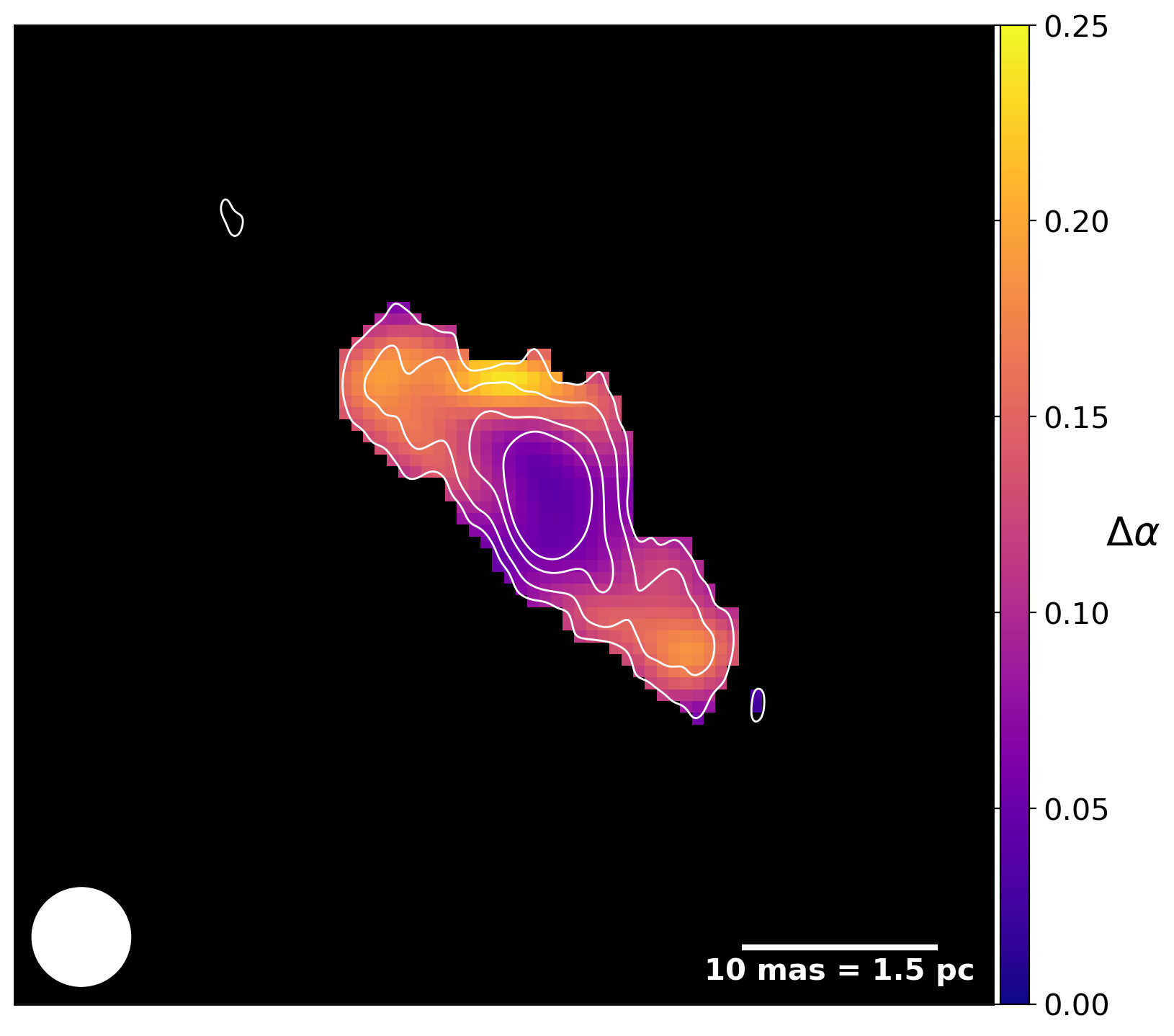}
    \caption{VLBA observation of the central continuum source of NGC~5044. \textit{Top left:} Combined image reconstructed at $\SI{6.7}{GHz}$ with contours at $-3$ (dashed), 3, 4, 8, 16, 32, 64, and 256 $\sigma$ levels, where $\sigma = \SI{16}{\mu Jy\,bm^{-1}}$. 
    \textit{Top right:} Reconstructed images at the two observing frequencies 5 and 8.5\,GHz.
    \textit{Bottom:} Spectral index map (left) with contours at 5, 15, 50, 100 $\sigma$, and the corresponding error map (right). }
    \label{fig:vlba}
\end{figure*}

After splitting the target scans with derived bandpass and gain calibrations applied, we proceed with further imaging and calibration tasks in the Common Astronomy Software Applications package  (CASA; \citealp{McMullin2007-ed}) Version 5.7.0. 
Before manual inspection, we used the rflag algorithm within the CASA task flagdata to remove residual RFI from our datasets. Imaging was performed using tclean with the mtmfs deconvolver algorithm (\citealp{Rau2011-gx}), Briggs weighting scheme (robust $=1$), and cleaning done on scales 0 and 10, so we are able to map extended emission more accurately. 
The images reconstructed from just one band (C or X) show the same structures (see Fig. \ref{fig:vlba} top right panel), therefore, we combine the datasets to one single image at the effective frequency $\SI{6.7}{GHz}$ (Fig. \ref{fig:vlba} top left panel).
Since we image all observations (both bands) at the same time to reach the lowest noise, we use two Taylor terms for the frequency dependence. After several phase-only self-calibrations, we also run an amplitude and phase self-calibration step at the end, which greatly improves our rms (about a factor of 4). Since amplitude self-calibrations can alter the flux of sources, we rescale the whole image so that the integrated flux of our entire source reflects the value before the amplitude self-calibration. Between each self-calibration step we manually identified additional antenna-based RFI and flagged it.

The combined imaging of both frequencies with two Taylor terms provides also an image of the spectral index. We use this as our spectral index map after some smoothing (5\,mas).

\subsection{JCMT}
\label{ch:jcmt_data}
The Submillimetre Common-User Bolometer Array 2 (SCUBA-2) is a bolometer camera operating at 450 and $\SI{850}{\mu m}$ on the James Clerk Maxwell Telescope (JCMT). These observing wavelengths extend our existing datasets to the highest frequencies above ALMA CO(2-1) observations, and help us to charactize the overall source spectrum of NGC~5044. NGC~5044 has been observed several times in various projects (see Tab. \ref{tab:jcmt}), from which obtained the reduced datasets from the Canadian Astronomy Data Centre (CADC).
\begin{deluxetable}{ccc}
 \tablecaption{JCMT Observations of NGC~5044\label{tab:jcmt}}
 \tablehead{
 \colhead{Project} &
 \colhead{Date} &
 \colhead{Exposure time}\\
 \colhead{} & 
 \colhead{} & 
 \colhead{850/450$\si{\mu m}$}
 }
\startdata
M13AU38 & 2013-07-03 & 2.2/0.6min\\
S14AU03 & 2014-06-08 & 1.1/0.3min\\
S14BU03 & 2015-01-25 & 1.1/0.3min\\
M15AI70 & 2015-04-27 & 1.2/0.3min\\
        & 2015-06-17 & 1.5/0.4min\\
M15BI025 & 2015-12-17 & 2.9/0.7min\\
         & 2015-12-25 & 2.9/0.7min\\
         & 2016-01-12 & 2.9/0.7min\\
M16AP083 & 2016-02-02 & 2.9/0.7min\\
         & 2016-04-17 & 2.8/0.7min\\
         & 2016-04-28 & 2.8/0.7min\\
\enddata
\end{deluxetable}

Further processing of the data products was done with the Starlink Software Package Version 2018A (\citealp{Currie2014-xh}). 
An overall source flux was estimated from a mosaic image (MOSAIC\_JCMT\_IMAGES) after removing regions outside a radius of $\SI{2}{\arcmin}$ (CROP\_SCUBA2\_IMAGES). To determine the flux, we used the beamfit task with the FWHM free to vary. The FWHMs of the fitted beams ($\SI{13.4}{\arcsec}$ and $\SI{9.8}{\arcsec}$, for 850 and $\SI{450}{\mu m}$, respectively) are close to the SCUBA-2 PSF (\citealp{Dempsey2013-ya}).
To determine the flux of the individual observation nights, we fixed the FWHM to the values in \cite{Dempsey2013-ya}, due to insufficient signal to noise.

\subsection{ATCA}
\label{ch:atca_data}
The Australia Telescope Compact Array (ATCA) is operated by CSIRO, and comprises six $\SI{22}{m}$ antennas, which can be moved along a railway track to reach a maximum east-west baseline of $\SI{6}{km}$.
NGC~5044 was observed with ATCA in three configurations between 2008 and 2015 (project C1958, PI A. Edge). 
On October 29/30, 2008,  NGC~5044 was observed for 46 min in the 6A configuration using a $\SI{128}{MHz}$ bandwidth centered on 4.8 and $\SI{8.6}{GHz}$, very close to our VLBA observing frequencies. 
On September 21/22, 2013, it was observed for 25 min in the more compact configuration H214 using a $\SI{2048}{MHz}$ bandwidth receiver centered on 19, 25, 32, and $\SI{38}{GHz}$. Finally, on February 23 2015, NGC~5044 was observed for 8 minutes with $\SI{2048}{MHz}$ bandwidth centered on 19, 24, 33, and $\SI{37}{GHz}$ in the 750D configuration.
We discarded the 2013 observations due to the lack of a stable flux calibration observation, and the 2015 observation due to the short on-target time.
The 2008 dataset was downloaded from the Australia Telescope Online archive\footnote{\url{http://atoa.atnf.csiro.au}}.
We used the CASA software package for the ATCA interferometric data analysis, which supports the native ATCA data format (importatca). We proceeded to flag edge channels, and used 1934-638 as the primary flux and bandpass calibrator. The flux model was chosen from the ATCA Calibrator Database v3\footnote{\url{https://www.narrabri.atnf.csiro.au/calibrators/}} by selecting the flux and spectral slope near the observing frequency and time.
The source 1245-197 was used as a phase calibrator, and fluxes have been bootstrapped to the scans of NGC~5044. 
Due to the poorly filled uv plane we could not create an image with the tclean task. However, since we are interested only in the total source flux, we used the uvmodelfit task to directly fit a point source model, with the position and flux as free parameters, to the calibrated visibilities. 

\section{Results}
\label{ch:results}
In the following, we describe our findings starting with the source structure and spectral properties in the VLBA observations. Our determination of the SED of the central radio source in NGC~5044 over more than three orders of magnitude allows us to fit physical models to the radio emission. Finally, the time variability of the JCMT and ALMA observations allows to check the VLBA constraints on the source extent.
\subsection{VLBA observations}
\label{ch:vlba_result}
With our combined image of all four observations (two at each frequency), we are able to reach a noise level of $\SI{21}{\mu Jy\,bm^{-1}}$ and a resolution of $4.5 \times \SI{2.5}{mas}$ ($0.68 \times \SI{0.38}{pc}$, see Figs. \ref{fig:cavities}d and  \ref{fig:vlba}). The noise is very uniform across our image, and the sufficiently filled uv-space helps us to map fainter, extended emission. We only see a few negative peaks at the $3 \sigma$ level around the source (see the dashed contours in Fig. \ref{fig:vlba}), indicating that the clean algorithm is able to model the source structure, and we are not affected by significant RFI.

The continuum source in NGC~5044 can be structured into a compact core in the center, and short jets to the NE and SW (Fig. \ref{fig:vlba}). The position of the core is
R.A.~$13^{\rm h}\,15^{\rm m}\,23.9613^{\rm s}$  \,Dec.~$\ang{-16;23;7.549}$ with position accuracy $\SI{1}{mas}$ (dominated by the accuracy of the phase reference source). This position is consistent with the reference from WISE measurements (\citealp{Cutri2013-qd}).
The length of the NW jet is $\SI{15}{mas} = \SI{2.25}{pc}$ and SW jet is $\SI{16}{mas} = \SI{2.4}{pc}$. The jets are oriented $\SI{41}{\degree}$ from N to E, so they are not extended along the major axis of the synthesized beam, which is in north-south direction. The integrated flux at $\SI{6.7}{GHz}$ is $\SI{22.2(5)}{mJy}$, consistent with measurements from ATCA (at 4.8 and $\SI{8.6}{GHz}$), and the extrapolated flux from the VLASS survey at $\SI{3}{GHz}$ (more details in Section \ref{ch:SED}). Therefore, we assume that our VLBA observations detect all emission from the central continuum source in NGC~5044, and no radio emission is found on larger scales than the maximum resolvable scale of the VLBA (about $\SI{50}{mas}$).

We also map only the compact emission by selecting baselines larger than $\SI{35}{M\lambda}$, where we find a total emission of $\SI{11(1)}{mJy}$, located in the center. Removing this compact emission from the visibilities using the uvsub task in CASA, we can map the extended emission component, which has a flux of $\SI{11(1)}{mJy}$.

We find peaks of emission along the jets (Fig. \ref{fig:vlba} top right), which possibly resemble hotspots: At 9 and $\SI{15}{mas}$ along the SW jet, and at $\SI{7.5}{mas}$ along the NE jet. However, the emission of these ``hotspots'' is not dominant over the central core.
Comparing the flux of the two jets, it turns out that the SW jet is only about 35\% brighter than the NE one, indicating that the jet orientation is close to the plane of sky (more details in Section \ref{ch:jet_direction}).

Since we have observations at two frequencies, $\SI{5}{GHz}$ and $\SI{8.5}{GHz}$, we can also analyze the spectral properties of the radio emission. The overall spectral index including both, extended and compact emission of the whole source is $\alpha = \num{-0.09(8)}$, making this an almost flat spectrum source. The compact emission, which is concentrated in the core, has a positive index of $\alpha_{\rm compact} = \num{0.18(10)}$, while the extended emission is steeper, $\alpha_{\rm extended} = \num{-0.38(11)}$.
The NE jet, excluding emission from the core, has a spectral slope of $\alpha_{\rm NE} = \num{-0.62(19)}$, while the SW jet has $\alpha_{\rm SW} = \num{-0.23(11)}$.

A more detailed spectral index map is shown in Fig. \ref{fig:vlba} (bottom panel). The spectral structure described in the paragraph above is also visible here: A core with a positive slope, and negative slope jets, while the NE jet is slightly steeper than the SW jet. We notice that the point of largest slope is not exactly in the center of the core, where the emission is brightest, but slightly offset to the NE. 
It is unclear whether this separation is real, or an artifact, since the offset is about one beam, and the uncertainty map does not indicate highly uncertain slopes in this particular region. 

\subsection{The spectral energy distribution from low frequency radio to sub-mm regime}
\label{ch:SED}

In Section \ref{ch:vlba_result}, we have seen that the structure of the central radio continuum source in NGC~5044 can be decomposed in a compact central source with a positive spectral index (flux increasing with frequency), and two jets in NE-SW orientation with a negative spectral index. As seen already with ALMA (\citealp{Schellenberger2020-vl,David2014-jn}) the total flux of the continuum source keeps rising with increasing frequency. Understanding the spectral behavior is important to draw conclusions on the current state of the AGN, and its accretion history. Since the X-ray data shows us the outburst history of the AGN, we can link it with the current state of AGN and the accretion history to make tight constraints on the feedback model for this nearby galaxy group.

\begin{deluxetable}{ccccc}[t]
 \tablecaption{Total source fluxes for NGC~5044}
 \tablehead{
 \colhead{Instrument} &
 \colhead{$\nu$} &
 \colhead{Flux} &
 \colhead{Year} &
 \colhead{Comment} \\
 \colhead{} &
 \colhead{GHz} &
 \colhead{mJy} &
 \colhead{} &
 \colhead{} 
 }
 \decimals
\startdata
GMRT & 0.240 & $\num{11.8(50)}$ & 2010 & $\geq 15k\lambda$\\
GMRT & 0.325 & $\num{18.6(19)}$ & 2010 & $\geq 15k\lambda$\\
GMRT & 0.610 & $\num{25.9(7)}$ & 2010 & $\geq 15k\lambda$\\
VLA & 1.5 & $\num{27.5(5)}$ & 2015 & $\geq 15k\lambda$\\
VLASS & 3.0 & $\num{24.1(5)}$ & 2019 & \\
ATCA & 4.8 & $\num{25.6(35)}$ & 2008 & \\
VLBA & 6.7 & $\num{22.2(5)}$ & 2020 & \\
VLBA & 6.7 & $\num{11.4(10)}$ & 2020 & jets\\
VLBA & 6.7 & $\num{11.1(10)}$ & 2020 & core\\
ATCA & 8.6 & $\num{22.2(18)}$ & 2008 & \\
ALMA & 100.3 & $\num{30.9(15)}$ & 2018 & \\
ALMA & 114.2 & $\num{32.3(15)}$ & 2018 & \\
ALMA & 243.3 & $\num{45.4(20)}$ & 2017 & \\
ALMA & 230 & $\num{45(20)}$ & 2016 & \\
JCMT & 352 & $\num{48(9)}$ & 2016 & \\
JCMT & 666 & $\num{36.8(48)}$ & 2016 & 
\enddata
\tablecomments{Fluxes of NGC~5044. \label{tab:fluxes}}
\end{deluxetable}

We start filling the SED of NGC~5044 with our VLBA observation at $\SI{6.7}{GHz}$, where we have not only a continuum flux, but also know the spectral behavior, and are able to separate the two main emission components (jets and compact core). As mentioned before, we also have fluxes at high frequencies from ALMA CO measurements. We use line-free spectral windows, intended for the continuum subtraction, to determine the flux of the continuum source at frequencies around $\SI{110}{GHz}$, the CO(1-0) line observation, and $\SI{230}{GHz}$, the CO(2-1) line observation. We suspect the continuum flux at $\SI{110}{GHz}$  given in \cite{Rose2019-au} of $\SI{14.6}{mJy}$ to be underestimated, since we find in our inspection of the same dataset a much higher flux in all spectral windows slightly above $\SI{30}{mJy}$. This is consistent with observations of the Combined Array for Research in Millimeter-wave Astronomy (CARMA, project cf0076, PI Werner, observed in July 2014), where we ran a basic analysis to determine the continuum flux ($\SI{29.4(18)}{mJy}$ at $\SI{108}{GHz}$ with indications for a positive spectral slope over the $\SI{10}{GHz}$ bandwidth). 

For the lowest frequencies, below $\SI{1}{GHz}$, fluxes have been published by \cite{Giacintucci2011-ru} and \cite{David2009-hn} showing a continuous increasing trend toward the lowest frequencies. However, the spatial resolution at these frequencies is coarse and extended emission around the AGN (e.g., from the inner cavities) might be blended with the central continuum source. Therefore, we use the reduced GMRT datasets at 235, 320, and $\SI{610}{MHz}$ (from \citealp{Giacintucci2011-ru}), and rerun the clean algorithm by selecting only baselines between 15 and $\SI{20}{k\lambda}$. This is the highest range for all three frequencies in common, to still get reasonable images.
We notice a clear drop in flux density toward the low frequencies, possibly caused by self absorption. 

In the range of 1 to $\SI{10}{GHz}$, we use the 2015 JVLA observation in L band (project 15A-243, in BnA configuration). We ran the standard VLA pipeline on this dataset and use the spectral windows without significant RFI and without the HI line, for continuum imaging (using the same uv-range as above) of the continuum source. We measured a source flux of $\SI{27.5(5)}{mJy}$ at $\SI{1.5}{GHz}$.

We also used the data products of the VLA Sky Survey (VLASS, \citealp{Lacy2020-kg}) at $\SI{3}{GHz}$ taken in 2019.  
We chose to work with the science-ready images provided by the VLASS archive, since the resolution of $\SI{2.5}{arcsec}$ is sufficient to avoid confusion with any nearby source.

The data reductions of the ATCA and JCMT/SCUBA-2 data are described in Sections \ref{ch:jcmt_data} and \ref{ch:atca_data}.
We note that there are more radio datasets available, that could be added here, for example the Parkes-Tidbinbilla Interferometer (PTI, \citealp{Slee1994-he}), Parkes single dish observations (\citealp{Disney1977-ue,Ekers1978-do}), historical VLA observations (\citealp{Giacintucci2011-ru}), or the $\SI{1.4}{GHz}$ NRAO VLA Sky Survey (NVSS). However, we decided to not include these data, since some of them are short and in some cases have too low resolution. Moreover, we describe in Section \ref{ch:variability} the time variability of the high frequency component, which disqualifies any older data from this SED fit. All our datasets have been observed after 2010, except the ATCA 5 and $\SI{8.5}{GHz}$ observations were taken in 2008, but are consistent with the fluxes from our recent VLBA observations. Also the low frequency GMRT observations were taken before 2010, but the low frequency component is not expected to be highly time variable, as we discuss below.

\begin{figure}[tp]
    \centering
    \includegraphics[width=0.5\textwidth]{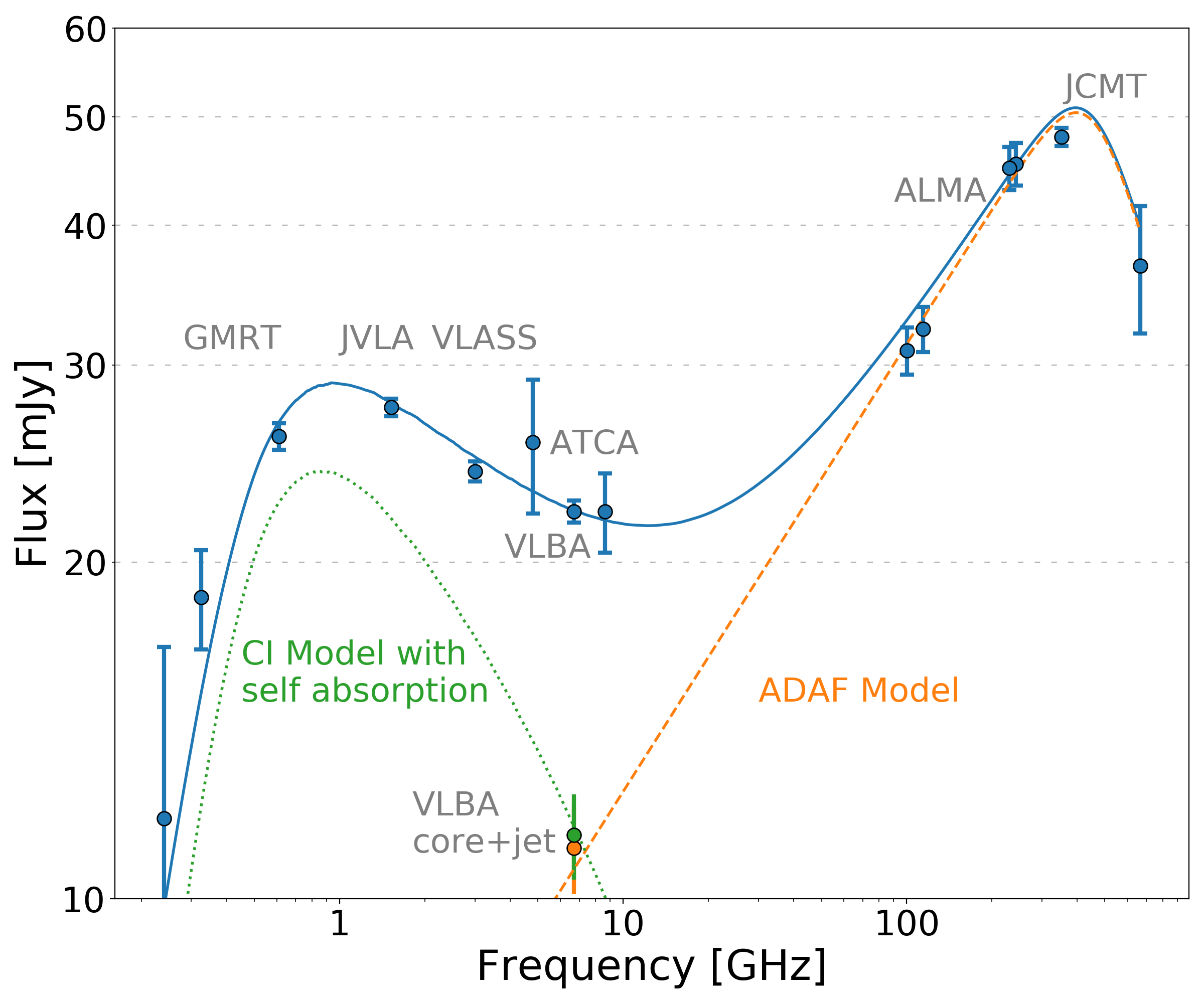}
    \caption{Spectral energy distribution of the central continuum source in NGC~5044. The green and orange datapoints illustrate the flux from the jets and core, respectively. The dotted green line corresponds to a CI model with injection index $\alpha_{\rm inj}$ left free to vary, and self absorption at the low frequency end. The dashed orange line is emission from advection dominated accretion. The solid line is the sum of the two models.}
    \label{fig:sed}
\end{figure}

The fluxes used for our SED fitting are listed in Tab. \ref{tab:fluxes} and shown in Fig. \ref{fig:sed}. We interpret the low frequency emission, with the peak around $\SI{1}{GHz}$, as emission from the jets/lobes that are still powered by the central AGN, to which we fit a Continous Injection Model (CI, \citealp{Pacholczyk1970-lq,Jaffe1973-cs,Harwood2015-do,Turner2018-qj}) and an optically thick self absorption component with a spectral index of $5/2$. The high frequency component comes from the core (as seen in the VLBA observation), so it is likely emission from the AGN itself by the accretion of matter. Indications for this accretion process were seen already as CO absorption lines in the spectra of NGC~5044, which indicated molecular clouds within the sphere of influence of the AGN (\citealp{Schellenberger2020-vl}). This emission is modeled by an advection dominated accretion flow (ADAF, \citealp{Narayan1995-oq,Narayan1995-zf,Mahadevan1997-rx,Yuan2014-td}), consisting of a rising part from synchrotron cooling, and a falling part in the sub-mm regime from Compton cooling processes. 
The discontinuities in the ADAF and CI models are slightly smoothed to avoid sharp edges.

\begin{deluxetable}{cDDc}[t]
 \tablecaption{Best fit parameters from SED fitting}
 \tablehead{
 \colhead{Parameter} &
 \multicolumn4c{Value} &
 \colhead{Unit}
 }
 \decimals
\startdata
$M$ & $2.2^{+1.2}_{-0.7}$ & $1.6^{+0.8}_{-0.6}$ & $\si{10^8\,M_\odot}$\\
$\dot M$ & $5.2^{+4}_{-2}$ & $7.2^{+4}_{-3}$ & $\si{10^{-2}\,M_\odot\,yr^{-1}}$ \\
$r$ & $34.4^{+18.5}_{-12.7}$ & $50.9^{+25.2}_{-16.7}$ & $\si{R_{\rm S}}$\\
$\delta$ & $1.3^{+0.4}_{-0.5}$ & $1.3^{+0.4}_{-0.4}$ & $\si{10^{-3}}$ \\ 
\tableline
$\nu_{\rm break}$ & $7.6^{+4.4}_{-3.0}$ & $33.0^{+9}_{-16}$ & $\si{GHz}$\\
$\alpha_{\rm inj}$ & $-0.31^{+0.05}_{-0.06}$ & $-0.5$ & \\
$\nu_{\rm SSA}$ & $0.42^{+0.03}_{-0.04}$ & $0.53^{+0.02}_{-0.02}$ & $\si{GHz}$\\
$J_0$ & $26.0^{+0.5}_{-0.6}$ & $28.2^{+0.3}_{-0.3}$ & $\si{mJy}$
\enddata
\tablecomments{The best fit parameters on the SED model from the MCMC. The first four rows correspond to the ADAF model and are the AGN mass, accretion rate,  the accretion radius in Schwarzschild radii, and the electron heating parameter. The lower four rows describe the CI model and are the break frequency, the characteristic frequency below which the synchrotron self absorption dominates, the normalization. The values in the third column show parameters when the injection index is frozen to $-0.5$ . \label{tab:sedfit}}
\end{deluxetable}

The best fit values of important free parameters of our combined absorbed CI and ADAF model are summarized in Table \ref{tab:sedfit}. The ADAF model contains two more parameters: The viscosity parameter $\alpha_\text{ADAF}$ and the gas to total pressure $\beta$. We set both, $\alpha_\text{ADAF}$ and $\beta$, with tight priors to the values in \cite{Mahadevan1997-rx}, $0.3$ and $0.5$, respectively. 
All parameters are constrainted in the MCMC and the parameter space is sufficiently explored, as tested by multiple chains starting from different regions in parameter space. 
We find a strong co-variance between the mass and accretion rate, the radius and the mass/accretion rate, and among all parameters of the CI model (injection index, normalization, break frequency, and characteristic absorption frequency). Derived quantities take into account this co-variance since MCMC chains are used.

The combined absorbed CI plus ADAF model fits our observations reasonably well, which is reflected in the $\chi^2_{\rm red} \approx 5$. This includes only statistical uncertainties, and no systematics due to inaccuracies of the flux scale, time variability of the source, or calibrators, which will lower the $\chi^2$. We also note, that the two model components show very good agreement with the VLBA fluxes for the jet and core (green and orange data points in Fig. \ref{fig:sed}).

We also tested the SED fitting by freezing the injection index to $-0.5$, for which the ADAF model parameters do not change significantly (see Table \ref{tab:sedfit}). The CI model, however, shows that the spectral break is almost unconstrained for a steeper injection index. This is mainly due to the poorly sampled frequency range between 10 and 100\,GHz. 
Leaving the injection index free to vary makes our default choice of model parameters due to the smaller $\chi^2_{\rm red}$. A shallower injection index than 0.5 has been observed in gigahertz peaked sources before (GPS; e.g., \citealp{Murgia2002-bf}).

\subsection{Variability}
\label{ch:variability}
\cite{Schellenberger2020-vl} describe a time variability of the continuum flux at $\SI{230}{GHz}$. This was only based on very few measurements from different telescopes (ALMA in various configurations and the ACA) over the last few years. At this frequency, the dominant emission mechanism is the advection dominated accretion. A time variability should be seen best at the peak frequency, around $\SI{400}{GHz}$. 

\begin{figure}[t]
    \centering
    \includegraphics[width=0.5\textwidth]{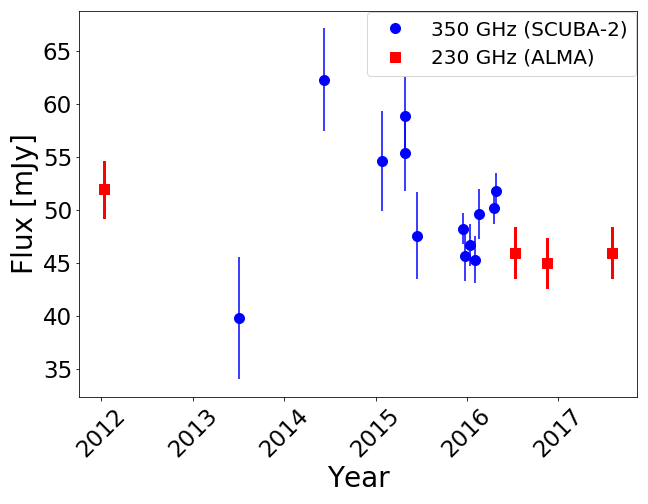}
    \caption{Time variability of the flux of NGC~5044 in the JCMT SCUBA-2 data at $\SI{353}{GHz}$ (blue) and the ALMA CO(2-1) data at $\SI{230}{GHz}$ (red).}
    \label{fig:variability}
\end{figure}

Since 2013 NGC~5044 was observed 12 times with SCUBA-2 at $\SI{353}{GHz}$. This allows us to create a lightcurve (see Fig. \ref{fig:variability}, blue datapoints). A linear trend is not obvious, although most flux measurements between 2014 and 2016 show a decreasing behavior, similar to ALMA (Fig. \ref{fig:variability} red datapoints). However, the earliest and latest SCUBA-2 measurements do not confirm this trend.
Assuming a periodicity by fitting a sinusoidal to the lightcurve results in a period $T=\SI{2.5(3)}{yr}$, which translates into a characteristic size of the source of $\SI{0.8(1)}{pc}$. This  is close to the upper limit of the core extent in the VLBA observation of $\SI{0.5}{pc}$. 
However, we emphasize that the current data is not sufficient to describe the temporal flux variation of the source completely. 

\section{Discussion}
\label{ch:discussion}
With the results presented in the previous section on the structure of the central continuum radio source in NGC~5044, and combined with the remarkable SED over more than three orders of magnitude, we are able to draw some conclusions about the ongoing AGN feedback in this galaxy group, and link the accretion process to the cooling of hot gas.

\subsection{Extent of the AGN}
The overall spectral index of the jets is $\num{-0.38(11)}$, which is close to the expected spectral behavior of jets ($-0.5$, \citealp{Carilli1991-fy,Komissarov1994-kv}). It is also consistent with the injection index of the CI model that we derive from the SED fit. We conclude that the AGN is currently active, and the small jets indicate a minor outburst, or the beginning of a major outburst.

In \cite{Schellenberger2020-vl} the authors found a spatially variable CO(2-1) absorption line with respect to the AGN continuum. Although the ALMA resolution did not resolve the AGN, the equivalent width of the two absorption lines seemed to vary across the beam, which led the authors to the hypothesis that the AGN might be extended below the ALMA resolution ($\approx \SI{60}{mas}$). Our VLBA observation now shows an extended central continuum source (about 30\,mas), which can be separated into a core and two jets. 
However, the extended part (the jets) has a negative spectral index, and the predicted flux of the jets at the frequency of the CO(2-1) line is less than 5\% of the total source flux. 
It seems unlikely that the extent of the jets causes the spatial variability of the previously observed absorption feature. Selecting only the longest baselines for the VLBA image gives a hint of an extent in NE-SW direction of the compact core. This extent is very small ($\SI{3}{mas}$), but might explain the absorption variability. Higher frequency VLBA observations are required to confirm the extent of the core with positive spectral index.

\subsection{Direction of the jets}
\label{ch:jet_direction}
Figure \ref{fig:vlba} shows the highest resolution image available for the central AGN in NGC~5044, which hosts a strong cooling flow. 
The clear jets show that the AGN feedback machine is currently active. The direction of the jets is almost perpendicular (in projection) to the location of the inner and intermediate cavity pairs, as well as the outer cavity with respect to the nucleus (see Fig. \ref{fig:cavities}). 

For NGC~5044 jet precession is not a likely explanation for the different direction of past outbursts.
The Lense-Thirring effect can cause a tilted disc orbit to precess for the case of a  spinning black hole (\citealp{Lense1918-zw,Nixon2013-zz}). However, timescales to change the direction of a jet in an AGN is typically more than $\SI{10}{Myr}$ when the accretion of matter is random but significant over a longer period of time. 
All the three cycles of past outburst are along the same orientation (NW-SE), so any precession must have started recently. The oldest cavities in the system (Fig \ref{fig:cavities} left) have an age of about $\SI{13}{Myr}$ (\citealp{David2017-ig}), while the innermost cavities must be significantly younger. \cite{David2017-ig} estimated their age from the sound crossing time $\sim \SI{1}{Myr}$. Precession is therefore unlikely to explain the change of the jet orientation. 

However, \cite{Babul2013-wa} demonstrate that the direction of jets might be changed by infalling clouds that form a thin accretion disk unaligned with the spin axis of the SMBH. This mechanism is proposed to solve the isotropic heat distribution provided by the jets. However, it is expected that a thin accretion disk also triggers the AGN to turn into a quasar, which is only observed very rarely in cool core clusters, possibly due to the small AGN duty cycle. In the case of NGC~5044 the mechanism offers a possible explanation for the misalignment of jets and cavities. However, the timescale since the last outburst ($\SI{1}{Myr}$) is short, and it only produced small cavities. The current jets are still weak, indicating the accretion of a smaller mass cloud, which is not able to change the jet direction significantly. 

Bent jets from AGNs are known, so it is possible that we are seeing here the effects of consistently and strongly bent jets. However, it is difficult to imagine an obstacle that would remain in the same place to cause the same degree of bending over multiple outburst cycles, which correspond to at least $\SI{10}{Myr}$.

If the jets were mainly aligned along the line of sight, the actual offset from the cavities could be much smaller than expected from the projected image.
The inclination to the line of sight can be estimated from the brightness ratio of the jets toward and away from the observer (\citealp{Lind1985-vr}).
This beaming effect also has been detected in the central AGNs of galaxy groups (e.g., \citealp{Giacintucci2008-wf}).
Although we do see the NE jet to be fainter than the SW, the difference is small, and does indicate that the jets are mostly, but not perfectly, in the plane of the sky. 
Based on the jet to counterjet flux ratio, we derive an inclination angle to the plane of the sky of $\theta = {9}_{-4}^{+10}\,\si{deg}$.
The spectral indices of the jets show a similar behavior - the NE jet is steeper - but the differences should be much larger for a jet orientation close to the line of sight.

A binary black hole system could also explain the different orientation of jets and cavities: Assuming that the central continuum source in NGC~5044 consists of two black holes in a binary system, we could be seeing the current jets from one of them, but the cavities were produced by an outburst of the other black hole. Systems like these, also with perpendicular jets to the X-ray cavities have been observed before (\citealp{Gitti2013-hn,Owen1985-yr,Rodriguez2006-cw}). However, even with VLBA resolution we do not see a second black hole, which could mean that either our resolution is not high enough, or the two black holes are along the line of sight or the second one is not accreting. A binary AGN system could be the remnant of a major merger (almost equal mass). 
\cite{Gastaldello2009-rr,David2011-du,OSullivan2013-ef} find a sloshing feature in the hot ICM and conclude that interaction with galaxy NGC~5054 perturbed the system about $\SI{1.2}{Gyr}$ ago. 
For triggering a double AGN the merger had perturbe also the central galaxy, NGC~5044, in the NGC~5044 group. However, this galaxy is known to have one of the least disturbed stellar components (\citealp{Tal2009-ik}) and a double AGN seems unlikely.
With the current data, the reason for the misalignment of the jets with the cavity direction remains unclear, and only higher frequency (i.e. higher resolution) VLBA data might be able to solve this puzzle.

\subsection{Advection dominated accretion}
\cite{Shakura1973-fw} described the accretion process of matter onto a black hole through the formation of a thin, optically thick disk. This theory predicts luminosities below the Eddington luminosity, otherwise a slim disk will develop. Since temperatures within the accretion flow are lower than the virial temperature, these models are classified as cold accretion models, and have been  applied to some AGN and Seyfert galaxies (e.g., \citealp{Koratkar1999-fx}). A hot accretion flow, on the other hand, with much higher temperatures will be optically thin, and lead to a two temperature plasma with hot ions and cooler electrons. Thermally stable solutions have been found by including advection, and allowing for heating by the accretion energy. The observational characteristic of hot accretion flows is a much lower radiation efficiency leading to lower luminosities. It has been shown that small accretion rates, typically $< 0.1 \frac{\dot M}{\dot M_\text{edd}}$, are not described by a thin disk model, but the ADAF as the most prominent example for hot accretion flows is favored (e.g. \citealp{Narayan1998-be,Yuan2003-ge}). Different regimes within hot accretion flows can be sub-classified by the mass accretion rate (e.g., clumpy ADAF accretion, \citealp{Wang2012-yy}), while in our case of a low accretion rate $\frac{\dot M}{\dot M_\text{edd}} \sim 0.01$ is best described by a pure ADAF. 

Our SED fit has shown that the ADAF model provides a good description of the accretion process in the center of this nearby galaxy group. The gas is transported toward the AGN by an advection dominated flow, and heated locally through the viscosity of the gas. 
Only a certain amount of this energy is transported inward through ions of the accreted gas, and the rest is transported to the electrons and radiated via synchrotron and inverse Compton emission (and bremsstrahlung at higher frequencies than observed here). 
It is expected that the energy fraction between electrons and ions is close to the fraction of electron to ion mass ($\sim 1/1800$). However, we find an electron heating parameter from our SED fitting, to be between ($1/570$ and $1/1200$), indicating a more efficient heating of electrons. At larger radii above $\num{1000}R_S$ the flow becomes cool and no significant radiation is emitted, while within this radius the electron temperature is constant and  the ion temperature is proportional to $1/r$ (\citealp{Narayan1995-zf}). A larger electron heating parameter might point towards emission not coming from the innermost region (within $3 R_S$), but further outside. 
Following the scaling relations by \cite{Mahadevan1997-rx} the expected electron temperature is $T_e \approx \SI{9.5e9}{K}$.

The standard luminosity of a thin disk can easily be derived from the accretion rate and the mass-energy-equivalence principle, $L_{\rm disk} = \eta_{\rm eff} \dot M c^2$,  
where $\eta_{\rm eff} = 0.1$ is the standard efficiency in converting mass into energy. For the case of NGC~5044, we expect a luminosity from the accretion disk of $\SI{3e44}{erg\,s^{-1}}$, which is much higher than observed. Furthermore, thin disk accretion would produce a black body spectrum with a steeper spectral index (within the Rayleigh-Jeans-regime) than we observe.
In the ADAF model, most of the heat energy is accreted onto the black hole and the predicted luminosity is lower, for NGC~5044, it is about $\SI{4e41}{erg\,s^{-1}}$, which is close to what is observed in the radio to sub-mm band (about $\SI{2e41}{erg\,s^{-1}}$). 
Remarkably good agreement of the black hole mass and the accretion rate is achieved through the SED fit. Both are in good agreement with values from independent methods. The ADAF black hole mass is only 3\% below the value derived from stellar dynamics ($\SI{2.27e8}{M_\odot}$, \citealp{David2009-hn}), and the statistical uncertainties are larger than this difference. The star formation rate derived from UV and IR observations (\citealp{Werner2014-vw}) is only an upper limit, which is just 40\% above the accretion rate derived from the ADAF model.  

There are fundamental limits to the ADAF model. For extremely low accretion rates no energy can be transferred to the electrons, which then do not emit synchrotron radiation. If the flow of matter is too large, the disk becomes optically thick resulting in a very different spectrum ($\propto \nu^{2.5}$). We do not observe such a steep spectral slope in NGC~5044. The critical accretion rate, below which advection occurs, is given by \cite{Mahadevan1997-rx}, which only depends on viscosity, $\dot M_{\rm crit} \sim \SI{0.12}{M_\odot\,yr^{-1}}$. NGC~5044 is clearly below this limit.

\subsection{Energy injection to the jets}
We model the low energy part of the NGC~5044 SED with a CI model. This model assumes a continuous injection of electrons than rather a single outburst, and is often used to describe the integrated emission of homogeneous lobes of extended radio galaxies. For simplicity, we assume a pitch angle scattering of the synchrotron electrons (JP version of the CI model), but this has a negligible effect (\citealp{Turner2018-qj}).

At the low energy end of the integrated spectrum, we see absorption likely caused by synchrotron self-absorption (SSA, see \citealp{Rybicki2008-dg,Blandford1979-oy}), characterized by a steep spectral slope of $+5/2$, independent of the slope of the electron energy distribution. Below the critical frequency $\nu_{\rm SSA}$ the optical depth is larger than unity and synchrotron photons are absorbed by the relativistic electrons.

Following the description for the peak synchrotron frequency in \cite{Kellermann1981-qo} and \cite{Tingay2003-mv} we can derive a source extent, i.e. the size of the base of the jets. However, this depends weakly on the magnetic field, which has been found to be on the order of several mG in a radio core (\citealp{OSullivan2009-im}). 
We can derive the magnetic field by minimizing the total energy content (see \citealp{Govoni2004-dx,Giacintucci2008-wf}), which yields $B = \SI{4.7}{mG}$.
With this we derive a source extent of about $\SI{0.7}{mas}$, which is just below our VLBA resolution. 
We think SSA might be a reasonable mechanism to cause the low energy turn-over, but it has also been pointed out that the magnetic field is likely not homogeneous and substructure along the jet will cause deviation from the ideal spectral shape.
Despite the fact that all our low frequency flux estimates come from one single radio interferometer and images have been extracted from the same spatial scales, it is likely that the larger beam size of GMRT includes some surrounding emission (e.g., the possible, faint radio emission of the inner cavities), which is unaccounted for in our SED model. 
Free free absorption by the hot ionized thermal plasma is unlikely since gas densities are orders of magnitude too low. 

From the spectral break in the CI model (see $\nu_{\rm break}$ in Table \ref{tab:sedfit}) one can calculate an age assuming a typical magnetic field.
Assuming $B=\SI{4.7}{mG}$ for the magnetic field strength in the jets, we derive an age of $1.7^{+0.8}_{-0.3}\si{kyr}$, which is significantly shorter than the age of the cavities. 
For the case of $\alpha_{\rm inj} = 0.5$ we find a slightly lower age of $0.8^{+0.6}_{-0.2}\si{kyr}$. This clearly shows that the launch of jets is not connected to the youngest cavities ($\sim \SI{1}{Myr}$ age), and thus marks a new, fourth visible AGN feedback cycle in NGC~5044. 

Considering frequencies below 10\,GHz, the SED resembles the spectrum of a GPS. Empirical correlations between the turnover frequency and the projected linear size have been found (e.g., \citealp{ODea1997-jv}). The linear size of the jets in NGC~5044 is about $\SI{4.5}{pc}$, while the turnover frequency is around $\SI{0.84}{GHz}$. This places NGC~5044 to the lower left in Fig. 3 from \cite{ODea1997-jv}, and given their sample of GPS and compact steep spectrum sources, we would expect a linear size of $\sim \SI{600}{pc}$ for NGC~5044. Clearly, we find that the radio source in NGC~5044 is very young and the jets have just been launched, which makes it challenging to compare this source with other sources in the literature.

X-ray cavities are formed by the buoyant rise of AGN jet inflated bubbles filled mainly with non-thermal plasma. They can be identified by a depression of the X-ray surface brightness and coincident radio synchrotron emission. 
Many studies have come up with scaling relations between the X-ray and radio characteristics of cavities (e.g., \citealp{Birzan2004-pw,Birzan2008-sl,Rafferty2006-qf,Cavagnolo2010-nx,OSullivan2011-xb}).
In NGC~5044 the innermost cavities are about $\SI{3}{arcsec}$ offset from the center and blend with the central radio source at GMRT resolution. 
Based on the X-ray properties of the surrounding medium around the inner cavities as stated in \cite{David2009-hn}, we estimate a cavity power for each of $P_{\rm cav} \approx \SI{9e39}{erg\,s^{-1}}$. 
If we assume a ratio between the cavity power (from the work expanding into ICM) and the radio luminosity of $\num{1000}$, we estimate a radio flux for both inner cavities of $\SI{2}{mJy}$ at $\SI{1.4}{GHz}$, and about $\SI{6}{mJy}$ at $\SI{325}{MHz}$, assuming a spectral index of $\num{-0.8}$. 
Using the scaling relations by \cite{OSullivan2011-xb}, which focuses on the cavities in galaxy groups and smaller systems, we estimate the radio flux in both inner cavities together to be less than $\SI{50}{\mu Jy}$ at $\SI{1.4}{GHz}$, and  $\SI{0.16}{mJy}$ at $\SI{325}{MHz}$. 
Even if we assume conservatively the higher flux estimates, the central AGN still dominates. 
Although we see in Fig. \ref{fig:sed} that the low frequency fluxes are slightly higher than the model predictions, it is unclear  if the difference is actually coming from the inner cavities. The VLA L-band image has the resolution and sensitivity to detect the flux from the inner cavities, but we do not find anything at the corresponding position. 
% past fluxes before correcting for uvrange at GMRT

\subsection{Time variability caused by infalling cold clouds?}
We see a time variability of the high frequency component in the SED in our ALMA and ACA data, but especially in the JCMT/SCUB2 data. 
If we interpret this as a periodic variability, we can derive an upper limit on the source extent, which is just below our VLBA resolution. 
Could this variability be triggered by infalling clouds onto the AGN? \cite{Schellenberger2020-vl} have shown that absorption features in the continuum spectrum correspond to small ($\sim \SI{e4}{M_\odot}$) molecular clouds most likely located within the sphere of influence of the black hole ($\sim \SI{17}{pc}$).
From the variability over 800 days, we can give a rough estimate of the ``extra'' energy emitted, and to what mass this corresponds. 
We notice a variation of about 15\% over 800 days, and we have seen that the total luminosity of the ADAF component is about $\SI{4e41}{erg\,s^{-1}}$. If this additional energy has been converted from mass with a standard efficiency of 0.5\%, we can translate this variability to a mass accretion of $\SI{5e-4}{M_\odot}$ over 800 days. This means that, in $\SI{50}{Myr}$, one cloud, of the size seen in the ALMA absorption spectrum, is accreted. 
More precise estimates on the variability can only be made through continuous monitoring at high frequencies.

\section{Summary}
\label{ch:summary}
We presented the results of our recent VLBA observations in C and X band of the radio continuum source in NGC~5044, one of the X-ray brightest cooling flow groups in the sky hosting the largest amounts of cold molecular gas. 
Due to its small distance, this system is the perfect example to study the nearby AGN feedback process from cooling of X-ray gas, formation of warm filaments, and condensation to cold molecular clouds, which feed the AGN. 
The remnants of older outbursts have been found in the past, but we have now the unique chance to look into the feedback process right at the ``edge'' of the AGN. 
Together with the subparsec-scale observations, we also used the radio and sub-mm SED to draw conclusions about the accretion process. The combination of these two measurements allows us to distinguish between emission from the jets (feedback process) and the accretion process. Our findings are:

\begin{itemize}
    \item a core-jet structure of radio continuum source indicating that a new outburst has just started, and that a new feedback cycle is building up. The size of the jets is small ($\SI{4.5}{pc}$) compared to other GPS, indicating the young age. Interestingly, the jets are likely in the plane of the sky ($9^{+10}_{-4}\,\si{deg}$), since we do not detect a significant beaming effect.
    \item The orientation of the jets is orthogonal to all previous outbursts (cavities). Possible explanations comprise a binary black hole system, precession of the jets, or strongly bent jets at larger distances. However, all these possibilities seem unlikely based on the current data.
    \item The spectral slope of jets and the core are very different: The core has a positive spectral index ($\alpha \approx 0.2$), which is also supported by the accretion model and higher frequency observations. The jets, as expected, have a negative spectral index ($\alpha \approx -0.4$) with the SW jet having a slightly shallower spectrum than the NE one.  
    \item The SED (from $\SI{240}{MHz}$ to $\SI{670}{GHz}$) is well fit by an absorbed CI plus ADAF model. This is one of the few examples, where the ADAF model has been successfully applied, and the derived accretion rate and SMBH mass are consistent with independent measurements. The equipartition magnetic field strength ($\SI{4.7}{mG}$) implies a very young age of the jets ($\leq\SI{2}{kyr}$). Since the innermost cavities are much older, we argue that we are now seeing the beginning of a new feedback cycle of the AGN.
    \item From the SCUBA-2 images, we find the ADAF component to be time variable on the order of months (which is supported by ALMA data). Assuming an oscillating variability, we derive a characteristic size of the core of about $\SI{0.8}{pc}$. If this brightness variation was caused by infalling molecular clouds, one would expect a cloud of about $\SI{e4}{M_\odot}$
    (which is seen in CO absorption) to fall onto the AGN every $\SI{50}{Myr}$.
\end{itemize}
Despite the findings presented here, some open questions remain, such as the size of the accretion disk and the origin of the time variability. 
To answer these questions a better sampling of the 10 to $\SI{100}{GHz}$ window is needed to precisely model the SED. 
VLBA observations at these frequencies will also allow to resolve the expected size of the emission from the accretion disk.
A monitoring of the AGN over several years at high frequency will help to better understand the time variability. 

\acknowledgments
G.S. acknowledges support through Chandra grant GO5-16126X. 
W.F.  and C.J. acknowledge support from the Smithsonian Institution and the Chandra High Resolution Camera Project through NASA contract NAS8-03060.
This study makes use of VLBA data. The National Radio Astronomy Observatory is a facility of the National Science Foundation operated under cooperative agreement by Associated Universities, Inc.

The James Clerk Maxwell Telescope is operated by the East Asian Observatory on behalf of The National Astronomical Observatory of Japan; Academia Sinica Institute of Astronomy and Astrophysics; the Korea Astronomy and Space Science Institute; Center for Astronomical Mega-Science (as well as the National Key R\&D Program of China with No. 2017YFA0402700). Additional funding support is provided by the Science and Technology Facilities Council of the United Kingdom and participating universities in the United Kingdom and Canada.
Additional funds for the construction of SCUBA-2 were provided by the Canada Foundation for Innovation.

The Australia Telescope Compact Array is part of the Australia Telescope National Facility which is funded by the Australian Government for operation as a National Facility managed by CSIRO.
We acknowledge the Gomeroi people as the traditional owners of the Observatory site.
\bibliographystyle{aasjournal}
\bibliography{Paperpile.bib}

\end{document}